\input{aipcheck}

\documentclass[
    ,final            
  ]
  {aipproc}

\layoutstyle{8x11single}

\begin{document}

\title{High $p_{T}$ inclusive hadron and photon spectra in pp and PbPb collisions}


\author{Sevil Salur on behalf of the CMS Collaboration}{
  address={Department of Physics, University of California, One Shields Avenue Davis, CA 95616-8677},
  email={salur@rutgers.physics.edu},
altaddress={Department of Physics $\&$ Astronomy, Rutgers, 136 Frelinghuysen Rd, Piscataway, NJ 08854},
}

\begin{abstract}

In this article, we report the inclusive charged hadron and isolated photon spectra at $\sqrt{s_{NN}}=2.76$ TeV PbPb collisions measured with the CMS detector at LHC. Charged particle momentum spectra are compared to a constructed reference measurement of proton-proton collisions at $\sqrt{s}=2.76$ TeV by using a combination of $x_T$ scaling and direct interpolation at fixed $p_{T}$.  Photon results are compared to next to leading order calculations at the same center of mass energy. Nuclear modification factors ($R_{AA}$) are calculated to study the properties of the medium.  While a large suppression is observed in the charged particle $R_{AA}$ at high transverse momentum, isolated photons show no modifications, revealing that PbPb collisions at  $\sqrt{s_{NN}}=2.76$ TeV produce a strongly interacting colored medium.

\end{abstract}

\maketitle


Hard probes, particles with high transverse momentum or large mass, can be used to probe the properties of the high energy density matter  created in ultra relativistic heavy ion collisions.  At the Relativistic Heavy Ion Collider (RHIC), one of the first observed signatures of `jet quenching' in heavy-ion collisions was the dramatic suppression of high-$p_T$ charged hadron spectra with respect to a scaling of the proton-proton spectra by the number of binary nucleon-nucleon collisions -- a scaling that is observed for direct photons, which escape the produced medium without interacting strongly~\cite{starsup,phenixphoton,brahms,phobos,phenix,star}. However, models with a wide range of parameters are able to describe the measurements of the nuclear modification factors at RHIC, leading to limited constraints upon the underlying physics \cite{bass,armesto,xin}.   Measurements from LHC will  constrain model parameters and will therefore elucidate the mechanism of jet quenching and the properties of the medium produced in heavy ion collisions.  

The CMS detector complex is utilized to collect data and reconstruct final state particles \cite{cms}.  A minimum-bias (MinBias) event selection (coincidence trigger signals in either side of the Beam Scintillator Counters or the Hadronic Forward Calorimeter) was required during the 2010 LHC run for PbPb collisions at $\sqrt{s_{NN}}=2.76$ TeV. The impact parameter of the two colliding nuclei (centrality) is determined with the MinBias sample using the total sum of energy signals from both Hadronic Forward Calorimeters \cite{cent}. The average number of binary nucleon-nucleon collisions ($< N_{coll} >$) is calculated from a Glauber model of the nuclear collision geometry \cite{glauber}.  The nuclear overlap function $ < T_{AA}  > $ is the ratio of  $< N_{coll}  >$ and the inelastic nucleon-nucleon cross section $\sigma_{NN} = (64 \pm 5)$ mb at $\sqrt{s_{NN}}=2.76$ TeV \cite{pdg}.

In addition to the MinBias event selection, optimized photon data taking requires electromagnetic cluster with $E_{T}  >  5$ GeV and software based high-level triggers with photon $E_{T}  >  15$  GeV.  To extend the statistical reach and to enhance the charged particle yields at high $p_{T}$, calorimeter-based high-$E_{T}$ jet triggers with uncorrected transverse energy thresholds of $E_{T}	= 35$ GeV and 50 GeV  are employed \cite{CMS-PAS-HIN-10-005}.  
Further details of isolated photon and inclusive charged hadron analyses can be found in \cite{CMS-PAS-HIN-11-002,CMS-PAS-HIN-10-005,yoon,kim}.

\begin{figure}\label{spectra}
  \resizebox{20pc}{!}{\includegraphics{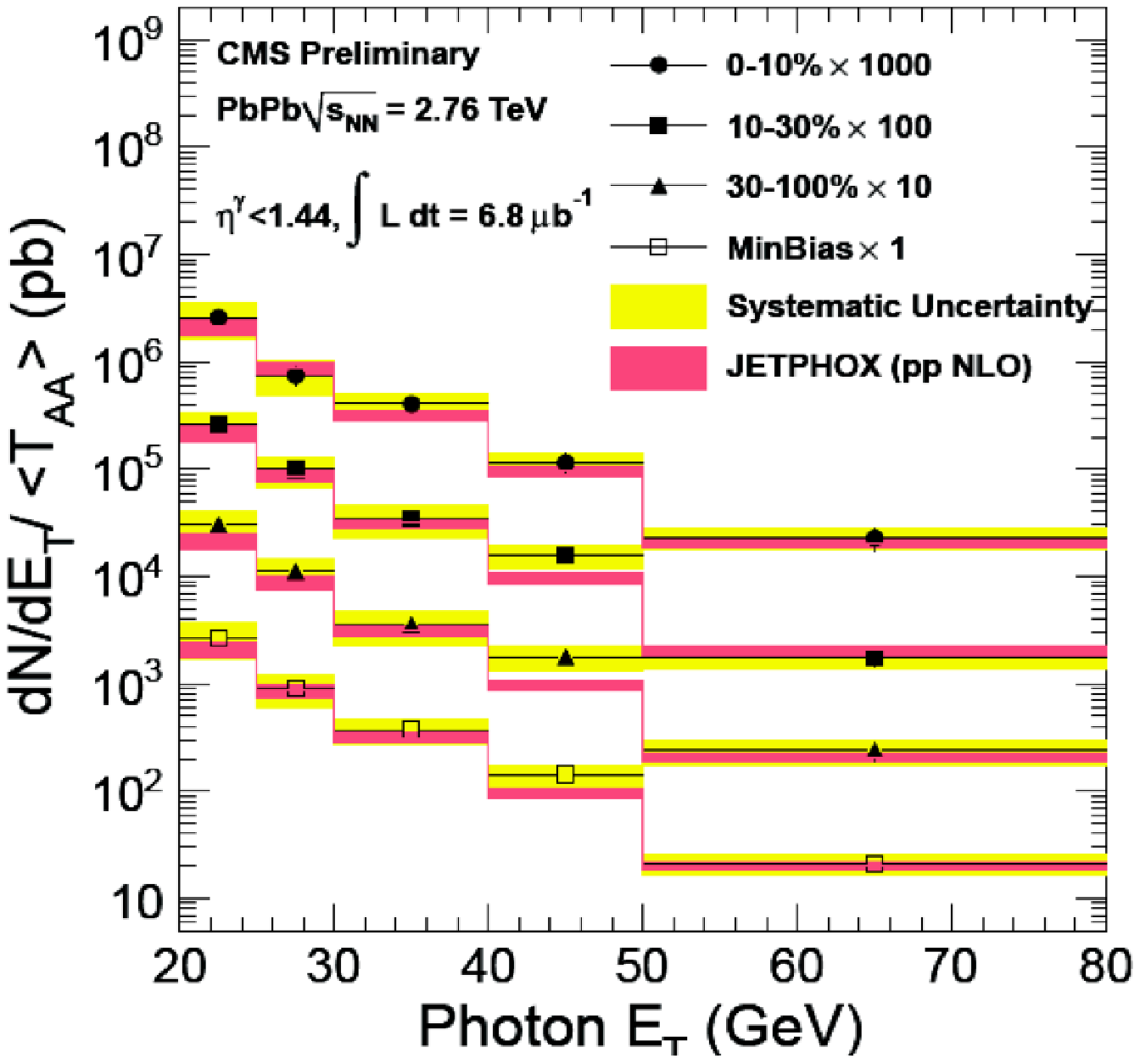}}\resizebox{18pc}{!}{\includegraphics{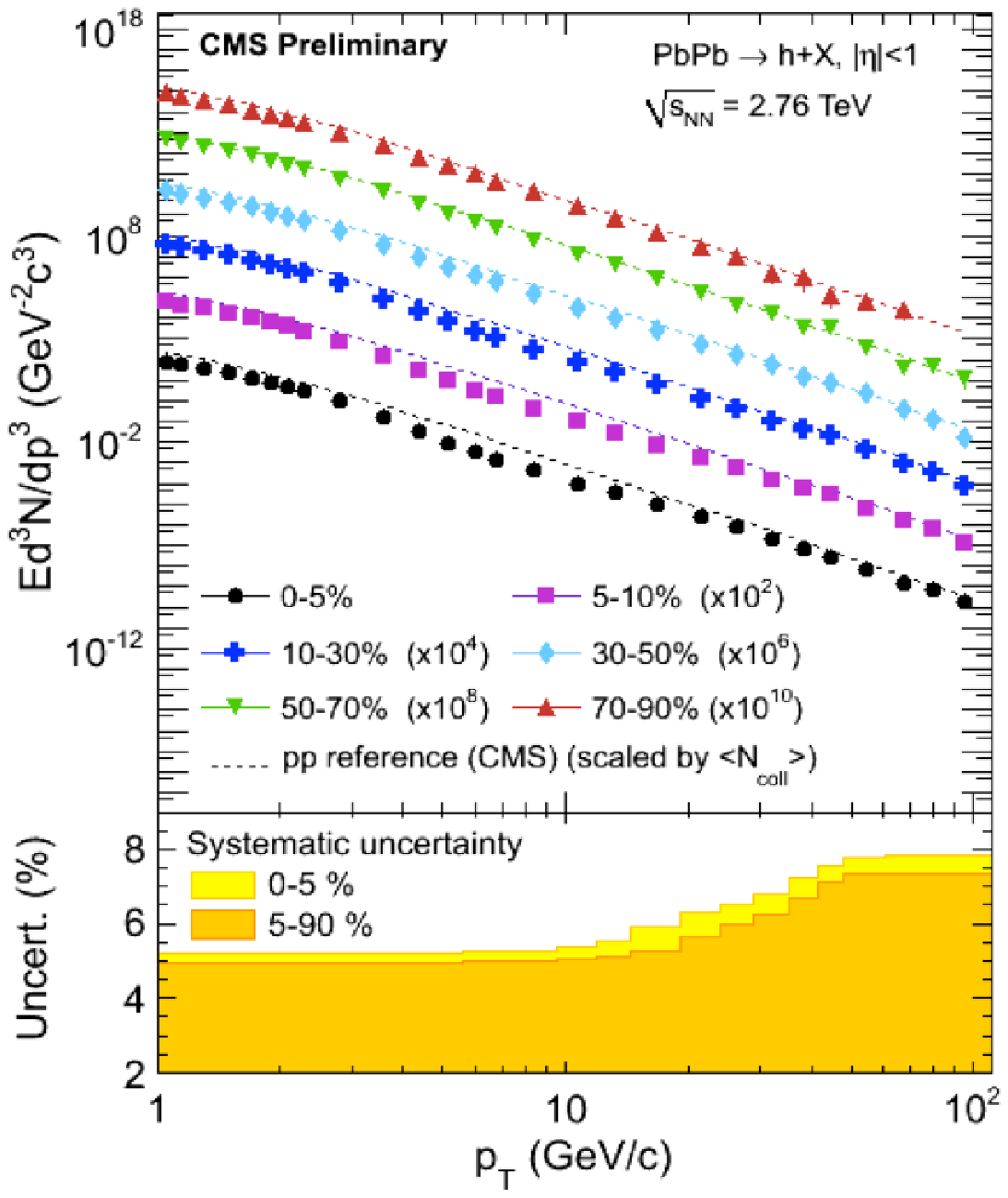}}
\caption{Left: Photon spectra in PbPb collisions are scaled by the inverse nuclear overlap function for the 0-10\%, 10-30\% and 30-100\% centrality bins. Right: Charged hadron spectra in PbPb collisions for the given centrality bins are compared with the scaled proton-proton reference.  }
\end{figure}

Isolated photon spectra for various centrality bins, shown in the left panel in Figure~\ref{spectra}, are scaled by the inverse nuclear overlap function and compared with JETPHOX calculations for expected proton-proton collision values \cite{Catani:2002ny}. Good agreement is observed with the calculated proton-proton predictions and scaled PbPb measurements within the given systematic uncertainties.   Inclusive charged hadron spectra for six different centralities are shown in the right panel in Figure~\ref{spectra}. These spectra are  compared to a nuclear overlap function scaled proton-proton reference at $\sqrt{s}=2.76$ TeV that is constructed by  using a combination of $x_T$ scaling and direct interpolation at fixed $p_{T}$ \cite{Chatrchyan:2011av}. While the most peripheral bins show good agreement with the scaled proton-proton reference, a deviation from this reference is observed in collisions with increasing centrality. Larger suppression at more central collisions is expected due to the longer average path lengths traversed by hard-scattered partons causing additional energy loss.

\begin{figure}[t!]\label{raa}
  \resizebox{18pc}{!}{\includegraphics{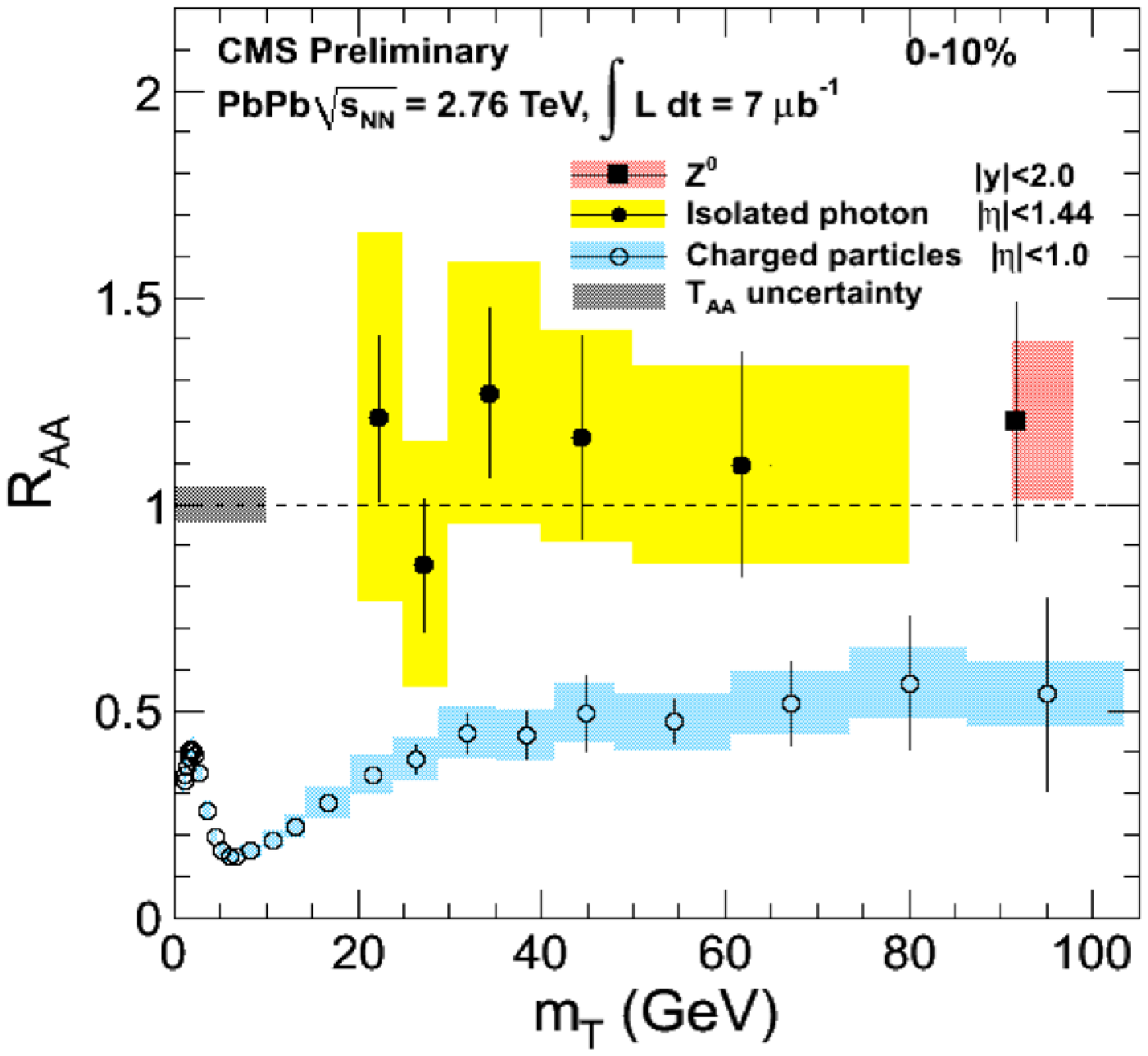}}\resizebox{18pc}{!}{\includegraphics{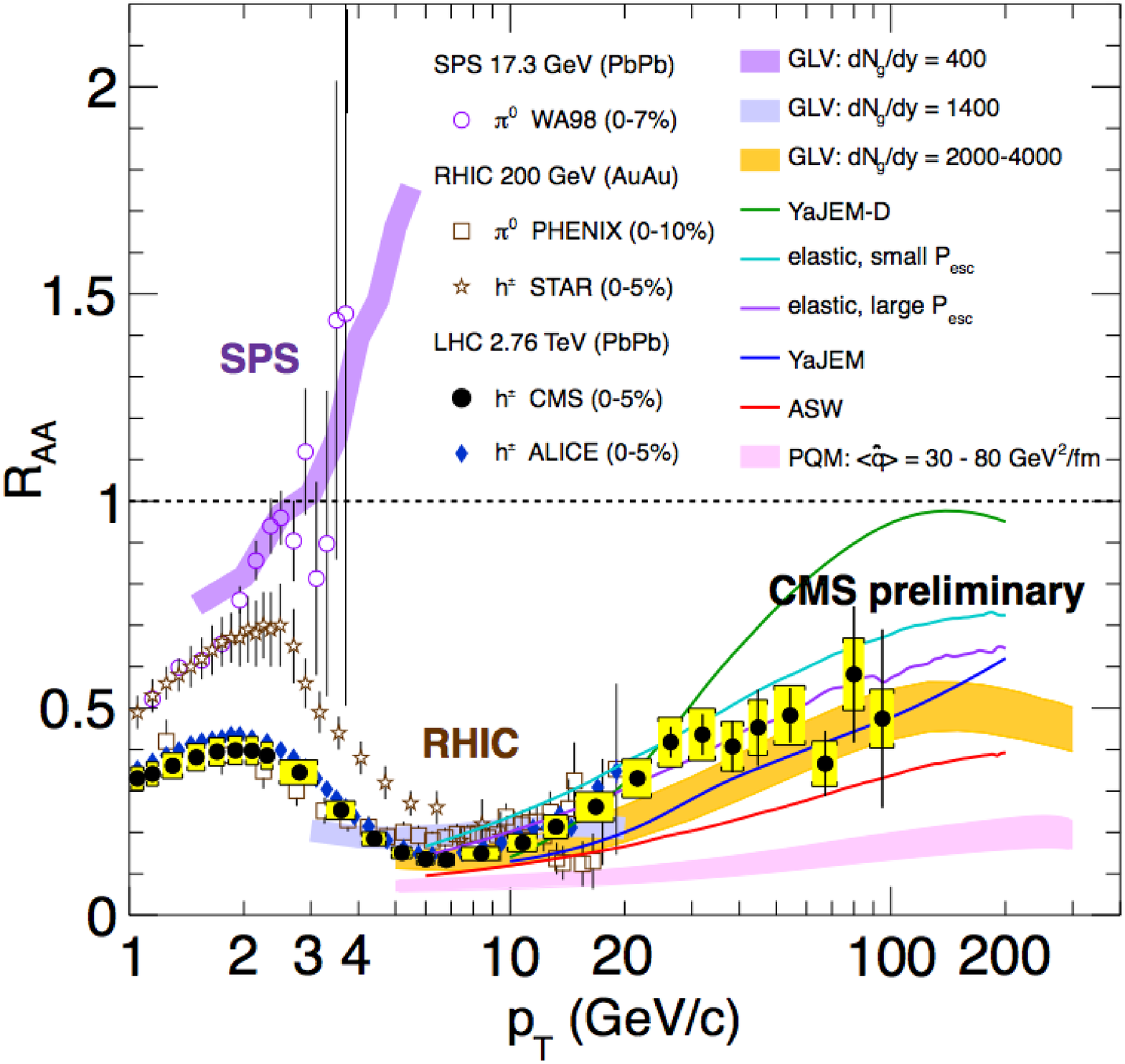}}
\caption{Left:The nuclear modification factors for charged hadrons, isolated photons, and Z bosons. Right: Model comparisons of nuclear modification factors measured at three different center-of-mass energies for neutral pions and charged hadrons. Boxes represent the systematic uncertainties while the error bars are statistical uncertainties. }
\end{figure}

To study high-$p_{T}$ particle suppression as done at lower energies at RHIC, a ratio of yields in PbPb to the proton-proton collision cross-section scaled by the nuclear overlap function, known as the nuclear modification factor $R_{AA}$, is calculated.  The left panel of Figure~\ref{raa} shows $R_{AA}$ measurements of isolated photons, charged particles and Z bosons for the 0-10\% most central collisions at $\sqrt{s_{NN}}=2.76$ TeV \cite{yoon,kim,jorge,yenjie}. Charged particles show a dramatic suppression in the 6-7 GeV/c transverse momentum range. At higher transverse momentum, the $R_{AA}$ value for charged hadrons rises, and it levels off above 40 GeV/c at a value of 0.5. Neither photons nor Z bosons show any modification with respect to theoretical next-to-leading order pQCD proton-proton collision cross sections scaled by the number of elementary nucleon-nucleon collisions since they weakly interact with the medium. This observation at unity confirms the validity of the Glauber scaling of PbPb collisions at $\sqrt{s_{NN}}=2.76$ TeV.

The center-of-mass energy evolution of nuclear modification factors from SPS to RHIC to LHC is also presented in the right panel of Figure~\ref{raa} \cite{Aggarwal:2001gn, Adare:2008qa, Adams:2003kv}. Although a larger charged particle suppression is observed at LHC than at RHIC, the similarity between the $\pi^0$ suppression observed at $\sqrt{s_{NN}}=200$ GeV by the PHENIX experiment and charged particle suppression at $\sqrt{s_{NN}}=2.76$ TeV is surprising.  The nuclear modification factors at large $p_{T}$ are also compared to theoretical predictions at the collision energy (elastic scattering energy-loss model with parametrized escape probability, ASW and YAJEM) and design LHC energy (PQM and GLV)   \cite{Vitev:2004bh, Renk:2011gj, Vitev:2002pf, Dainese:2004te}.    The magnitude of the rise for the nuclear modification factors varies significantly by models, even though they all predict a generally rising behavior.  

To summarize, the spectra of the inclusive charged hadrons and isolated photons are measured in PbPb collisions at $\sqrt{s_{NN}}=2.76$ TeV  in various centrality bins.  Nuclear modification factors are constructed with next-to-leading order perturbative QCD theoretical calculations for isolated photons and with a combination of $x_T$ scaling and direct interpolation at fixed $p_{T}$ utilizing measurements from various center-of-mass energies for inclusive charged hadrons for their proton-proton cross-sections. A large suppression  is observed for the inclusive charged hadron nuclear modification factor, providing evidence of jet quenching as seen at the lower collision energies at RHIC.  However, no modification is observed in isolated photons and Z bosons, confirming the validity of the Glauber scaling of PbPb collisions at $\sqrt{s_{NN}}=2.76$ TeV. These measurements of nuclear modification factors as a function of transverse momenta at LHC together with the measurements from lower energy collisions at RHIC will constrain the parameter space of models. This will illuminate the properties of the medium produced in heavy ion collisions through the jet quenching mechanism.

This material is based upon work supported by the National Science Foundation.




\bibliographystyle{aipproc}   

\bibliography{panic_salur}

\IfFileExists{panic_salur.bbl}{}
 {\typeout{}
  \typeout{******************************************}
  \typeout{** Please run "bibtex \jobname" to optain}
  \typeout{** the bibliography and then re-run LaTeX}
  \typeout{** twice to fix the references!}
  \typeout{******************************************}
  \typeout{}
 }

\end{document}